\documentclass[reprint,amsmath,amssymb,aps,prb,]{revtex4-2}

\usepackage{graphicx}
\usepackage{dcolumn}
\usepackage{bm}
\usepackage{subfigure}

\begin{document}
	\preprint{APS/123-QED}
	
	\title{Near coincidence of metal-insulator transition and quantum critical fluctuations: Electronic ground state and magnetic order in Fe$_{1-x}$Co$_{x}$Si}
	
	\author{J. Grefe$^1$, P. Herre$^1$, Y. Hilgers$^1$, F. Labbus$^1$, N. L\"uer-Epping$^1$, N. Radomski$^1$, M. A. C. de Melo$^{1,2}$, F. J. Litterst$^1$, D. Menzel$^1$, S. S\"ullow$^1$}

	\affiliation{$^1$Institut f\"ur Physik der Kondensierten Materie, TU Braunschweig, D-38106 Braunschweig, Germany\\
	$^2$ Departamento de Fisica, Universidade Estadual de Maring\'a, Maring\'a, PR, 87020-900, Brazil}
	
	\date{\today}
	
	\begin{abstract}
		We present a detailed study of the electronic and magnetic ground state properties of Fe$_{1-x}$Co$_{x}$Si using a combination of macroscopic and microscopic experimental techniques. From these experiments we quantitatively characterize the metal-insulator transition and magnetic/non-magnetic quantum phase transition occurring at low doping levels in Fe$_{1-x}$Co$_{x}$Si. From our study, we find a surprising closeness of the critical composition of the metal-insulator transition at $x_{\mathrm{MIT}} = 0.014$ and the quantum phase transition at $x_{\mathrm{LRO}} \sim 0.024-0.031$. This suggests that these effects are cooperative and depend on each other.
	\end{abstract}
	
	\pacs{Valid PACS appear here}
	\maketitle
	
	\section{\label{sec:level1}Introduction}
	
\begin{figure}[h]
    \includegraphics[width=1\linewidth]{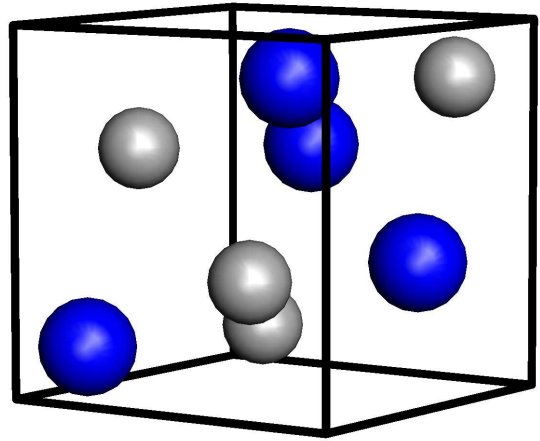}
		\caption{\label{structure} Cubic $B$20 crystal structure for FeSi, with blue and grey spheres representing Fe and Si, respectively (lattice parameter $a \sim 4.48$\,\AA ).}
\end{figure}

The class of $B$20 materials (Fe,Co,Mn)Si has been studied for decades, representing model compounds in various contexts of modern solid state physics. The materials crystallize in the cubic $B$20 crystal structure \cite{Pearson1958} (Fig. \ref{structure}). It is also labeled the {\it FeSi}-structure, as early on the compound was the most prominent representative of this crystallographic lattice that lacks inversion symmetry \cite{Foex1938}. Moreover, FeSi was the first material to attract attention with respect to its electronic and magnetic properties, with initial reports on a ''semiconductive and metallic'' ground state \cite{Wertheim1965,Shinoda1966} in the presence of an unusual magnetic behavior from ''correlated magnetic excitations'' \cite{Jaccarino1967}. MnSi, instead, was characterized as a ferromagnetic metal, while CoSi was reported as semimetallic diamagnet \cite{Shinoda1966}.

Subsequently, these observations were substantially refined. By now, MnSi has been established as a helimagnetic metal (for a review, see Pfleiderer et al. \cite{Pfleiderer2010}), where helimagnetism arises from the action of the Dzyaloshinskii-Moriya interaction \cite{DM} induced by the lack of inversion symmetry in the lattice. As result of the interplay of complex magnetic couplings and anisotropies a novel magnetic state, the {\it skyrmion lattice}, emerges in certain parameter ranges of the magnetic phase diagram \cite{Roessler2006}. For CoSi, recently the description was complemented by the realization that for a crystal structure lacking inversion symmetry in the presence of spin-orbit coupling it gives rise to new topological electronic states \cite{Bradlyn2016,Tang2017,Wang2020}.
	
Regarding FeSi, for a long time the central scientific issue was the nature of the small-gap semiconducting ground state \cite{Aeppli1992,Schlesinger1993}. It was proposed that this may be understood as signatures of a Kondo insulating state, {\it i.e.}, a semiconducting state arising as result of strong electronic correlations. Subsequent experimental tests of this concept have not produced clear-cut evidence in favor of this scenario \cite{Zur2007,Menzel2009}. Instead, it appears that single electron band structure modeling is sufficient to account for the observed electronic ground state. More recently, topological aspects of the band structure of FeSi have attracted attention \cite{Changdar2020}. In the present context, with respect to the electronic properties we will consider the bulk material FeSi as intrinsically gapped material, {\it i.e.}, an insulator in the traditional sense, which has metallic surface states \cite{Fang2018}. If these surface states reflect the character of a 3D topological insulator will not be addressed with our study. 
	
Another line of inquiry regarding these $B$20 compounds are alloying studies. With a full elemental solubility, alloying studies on Fe$_{1-x-y}$Co$_x$Mn$_y$Si allow to investigate both zero temperature (quantum phase) transitions of the electronic and magnetic ground states. Special focus lies upon the phase diagram of Fe$_{1-x}$Co$_x$Si, where a multitude of studies have been carried out in the course of 50 years of research \cite{Shinoda1972,Dubovtsev1972,Montano1975,Bloch1975,Kawarazaki1976,Beille1979,Beille1983,Watanabe1985,Motokawa1987,Povzner1989,Shimizu1990,Ishimoto1992,Lacerda1993,Mandrus1994,Chernikov1997,Aeppli1999,Manyala2000,Chattopadhyay2002a,Chattopadhyay2002b,Luo2003,Manyala2004,Menzel2004,Guevara2004,Onose2005,Mena2006,Zur2007,Grigoriev2007,Menzel2009,Munzer2010,Mazurenko2010,Yu2010,Forthaus2011,Patrin2011,Koralek2012,Shanmukharao2012,Milde2013,Schwarze2015,OuYang2015,Bauer2016,Zhang2016,Shanmukharao2018a,Shanmukharao2018b}. 

The general findings with respect to magnetic order are well-established \cite{Shinoda1972,Bloch1975,Kawarazaki1976,Beille1979,Beille1983,Watanabe1985,Shimizu1990,Ishimoto1992,Chernikov1997,Manyala2000,Chattopadhyay2002a,Chattopadhyay2002b,Onose2005,Munzer2010,Shanmukharao2012,Bauer2016,Zhang2016,Shanmukharao2018a,Shanmukharao2018b,Manyala2004,Grigoriev2007}: The series starts with the paramagnetic small gap insulator FeSi. Already alloying in the percentage range with Co closes the gap and induces the onset of long-range helimagnetic order below $T_{\mathrm{HM}}$, with a maximum ordering temperature in the range of a few 10\,K. Magnetic order is fully suppressed at $x = 0.8$, and with larger $x$ the series further transforms into the topological semimetal CoSi. In the magnetically ordered regime it is possible to identify field induced skyrmionic phases \cite{Munzer2010,Yu2010,Koralek2012,Milde2013,Schwarze2015,Bauer2016}. Since the parameter range of the formation of skyrmions in Fe$_{1-x}$Co$_x$Si is very different from that of MnSi, it has enriched the possibilities to quantitatively study the physics of skyrmionic spin textures.

\begin{figure}[h]
    \includegraphics[width=1\linewidth]{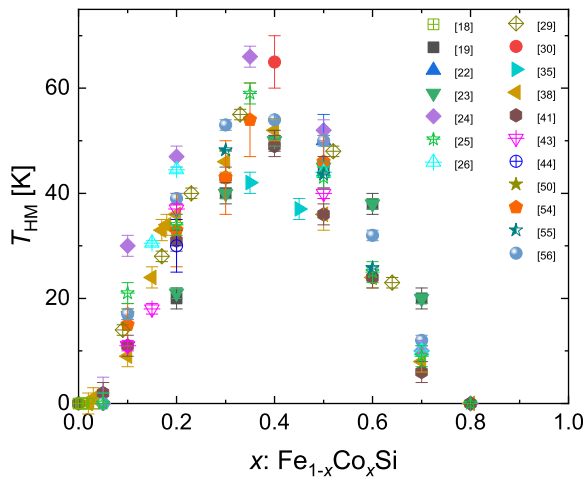}
		\caption{\label{THM} Helimagnetic ordering temperatures $T_{\mathrm{HM}}$ of Fe$_{1-x}$Co$_x$Si, as taken from literature (references listed in the legend); for details see text.}
\end{figure}

In detail, however, the magnetic and electronic phase diagram of Fe$_{1-x}$Co$_x$Si is far from well-established. To illustrate this point, in Fig. \ref{THM} we summarize the helimagnetic transition temperatures $T_{\mathrm{HM}}$ for samples of the series Fe$_{1-x}$Co$_x$Si reported in the literature. Here, the variations of the transition temperatures for a given composition exhibit a very large variation (even in the maximum of the $T_{\mathrm{HM}}$-dome by 30\%), which in other scientific contexts (high-$T_{\mathrm{C}}$ materials, quantum phase transitions etc.) would be considered unacceptable.

As pointed out by Bauer, Garst and Pfleiderer \cite{Bauer2016}, part of the problem are the different definitions and criteria chosen in literature to define the phase transition temperature $T_{\mathrm{HM}}$. In Ref. \cite{Bauer2016} this was illustrated by using different criteria on their selected samples Fe$_{1-x}$Co$_x$Si to extract $T_{\mathrm{HM}}$, leading to a large error bar in the determination of $T_{\mathrm{HM}}$ (up to $\pm 7$\,K). Still, Fig. \ref{THM} demonstrates that it does not fully account for the scatter of the data, and other effects need to be considered. These could be related to the use of poly- vs. single-crystalline samples, differing metallurgical treatment, inaccuracies in determining the correct stoichiometry (FeSi has a homogeneity range of formation), or - given that the literature reports span a range of 50 years - simple thermometry issues etc. in certain studies. At any rate, at this point a full, thorough and reproducible determination by well established techniques and criteria of the phase diagram of Fe$_{1-x}$Co$_x$Si is lacking. In particular, scientific topics with regard to the relationship of the metal-insulator transition (MIT) occurring in Fe$_{1-x}$Co$_x$Si at \cite{Chernikov1997,Manyala2000} $x \sim 0.005 - 0.018$ and the quantum phase transition (QPT) into a long-range magnetically ordered state at larger Co concentrations are simply not accessible with the published data. 

In this situation, we have set out to re-investigate the phase diagram of Fe$_{1-x}$Co$_x$Si. Our particular focus lies on the small-$x$ range, {\it i.e.}, $x \leq 0.15$, that is the range that encloses the MIT and the QPT. For our set of single-crystalline samples we perform a thorough characterization by various bulk experimental techniques accompanied by the microscopic technique M\"ossbauer spectroscopy. Taken together, we aim to shed light in particular on the physical phenomena occurring at small values $x$, that is the regime of magnetic quantum criticality and metal-insulator transition. 

\section{\label{sec:level2}Experimental details}

In the choice of our samples, we restrict ourselves entirely on single-crystalline specimens obtained by the Czochralski-method using a three-arc oven as described previously \cite{Zur2007,Menzel2009}. In the low-doping regime, between different samples, we choose particularly small variations of $x$ down to $0.01$, to accurately define the details of the magnetic phase diagram and electronic ground state properties. After growth, the samples have been oriented by means of Laue x-ray diffraction and bar-shaped samples have been cut along the cubic main axis from the crystals. 

\begin{figure}[h]
    \includegraphics[width=1\linewidth]{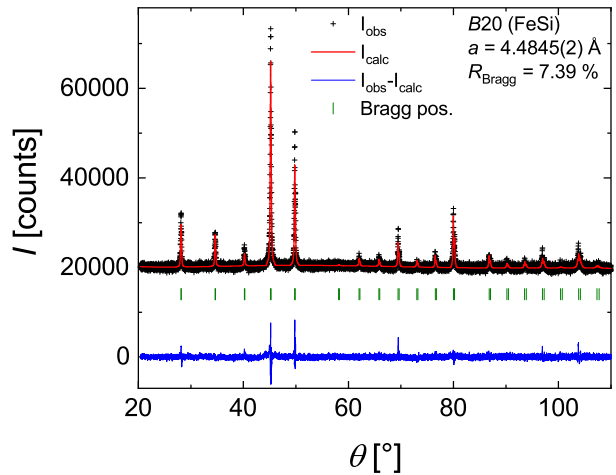}
		\caption{\label{lattice} Powder x-ray diffraction pattern as function of angle for FeSi. Rietveld refinement carried out using the $B20$ structure with lattice parameter indicated.}
\end{figure}

For each of the crystals some material has been ground to powder and checked by powder x-ray diffraction for phase homogeneity, crystal structure and lattice parameters. In the powder diffraction experiments no secondary phases have been detected and the crystal structure was verified as cubic $B$20 lattice. As an example, in Fig. \ref{lattice} we depict the x-ray diffraction pattern for FeSi, including a Rietveld refinement of the data. Results of similar quality are observed for the other samples, including some with larger $x$ to cover the full phase diagram. With the similarity in x-ray scattering cross sections of Fe and Co, an analysis of the actual composition of our specimens can not be carried out with a sufficiently high accuracy. Therefore, in the refinements we have used the nominal composition as fixed parameter.

From the x-ray analysis we obtain the evolution of the cubic lattice parameter as function of alloying which we list in Tab. \ref{fesi}. According to Vegard's law, with the lattice parameter of FeSi, $a = 4.4845$\,\AA , significantly larger than that of CoSi, $a = 4.4423$\,\AA, a (close to) linear shrinking of the lattice parameter would be expected with Co doping \cite{Wu2019}. Broadly speaking, this is observed experimentally in the alloying dependence of the lattice parameter. The other main free structural parameters in the $B20$ structure, the $x,y,z$ position of (Fe/Co) and Si, remain basically constant at $\sim 0.14$ (Fe/Co) and $\sim 0.84$ (Si) within experimental scatter.


\begin{table}
\caption{\label{fesi} Lattice and positional parameters in the $B$20 structure of Fe$_{1-x}$Co$_x$Si, derived from room temperature powder x-ray diffraction.}
\begin{center}
	\begin{tabular}{c|c|c|c}
$x$ & $a$ (\AA ) & Fe/Co & Si \\ \hline
0~ & ~4.4845(2)~ & ~0.134(1)~ & ~0.837(2) \\
0.01~ & ~4.4842(2)~ & ~0.137(2)~ & ~0.842(2)  \\
0.02~ & ~4.4867(2)~ & ~0.136(1)~ & ~0.843(2)  \\
0.03~ & ~4.4859(1)~ & ~0.135(2)~ & ~0.848(2)  \\	
0.04~ & ~4.4846(2)~ & ~0.138(1)~ & ~0.846(2)  \\
0.05~ & ~4.4851(3)~ & ~0.139(1)~ & ~0.837(2)  \\
0.06~ & ~4.4830(4)~ & ~0.137(7)~ & ~0.845(10)  \\
0.08~ & ~4.4816(10)~ & ~0.137(7)~ & ~0.843(8)  \\
0.15~ & ~4.4809(4)~ & ~0.138(6)~ & ~0.843(8)  \\
0.3~  & ~4.4756(1)~ & ~0.137(1)~ & ~0.841(1)  \\
0.55~ & ~4.4664(1)~ & ~0.141(1)~ & ~0.840(1)  \\ 
0.8~  & ~4.4561(1)~ & ~0.140(1)~ & ~0.836(1)  \\
1~    & ~4.4423(5)~ & ~0.145(8)~ & ~0.844(11) \\
	\end{tabular}
\end{center}		
\end{table}

To characterize the single crystals Fe$_{1-x}$Co$_x$Si regarding their electronic and magnetic properties we have carried out a standardized set of experiments. We report on the resistivity, magnetization, susceptibility and M\"ossbauer spectra, all in the $^4$He-temperature range, {\it i.e.}, 1.6 to 300\,K. For the resistivity we used a standard 4-probe {\it ac}-setup. Magnetization and susceptibility have been measured using a commercial SQUID system in fields up to 5\,T. 

M\"ossbauer experiments have been performed in a standard transmission geometry employing a 50\,mCi $^{57}$Co source in Rh-matrix. The samples used as absorbers obtained from several grinding and polishing runs of single-crystalline plates have been almost circular platelets (surface perpendicular to [100]) with maximum planar dimensions of 6\,mm diameter and thickness of about $50\mu$m. The gamma ray direction was along [100]. The M\"ossbauer velocity drive system was run in sinusoidal mode. The measurements were carried out in a liquid helium bath cryostat in under-pressure mode enabling experimental temperatures down to 1.7 K. The spectra were analyzed using the Mosswinn 4.0i software \cite{Mosswinn}. The spectra were fitted with two sites using the mixed magnetic and quadrupole static Hamiltonian (single crystal) theory, with the same quadrupole splitting $QS$, isomer shift $IS$ and the hyperfine magnetic field $B_{\mathrm{HF}}$ for both sites and differing only in the angles between $QS$ and $B_{\mathrm{HF}}$, between $QS$ and the gamma ray and $B_{\mathrm{HF}}$ and the gamma ray. The values of isomer shifts are given relative to $\alpha$-Fe at room temperature. 

\section{\label{sec:level3}Results}

Prior to a detailed investigation of the electronic and magnetic phase diagram of Fe$_{1-x}$Co$_x$Si, we need to establish the relevance of the metallic surface states for the interpretation of the experimental data. To this end, we utilize the argument put forth for topological insulators that the resistivity may be modeled as superposition of surface and volume electrical conduction \cite{Wolgast2013}. This observation implies that the relevance of surface conduction depends on the surface-to-volume ratio of a given sample \cite{Fang2018}. We prepared two specimens of our single crystal FeSi for resistivity measurements: first, a bar-shaped sample of dimensions $5 \times 1 \times 1$\,$\mathrm{mm}^3$, and secondly, a sample of similar length and width, but with a thickness polished down to $35 \mu$m, {\it i.e.}, with a surface-to-volume ratio increased by a factor of about 30. 

\begin{figure}[h]
    \includegraphics[width=1\linewidth]{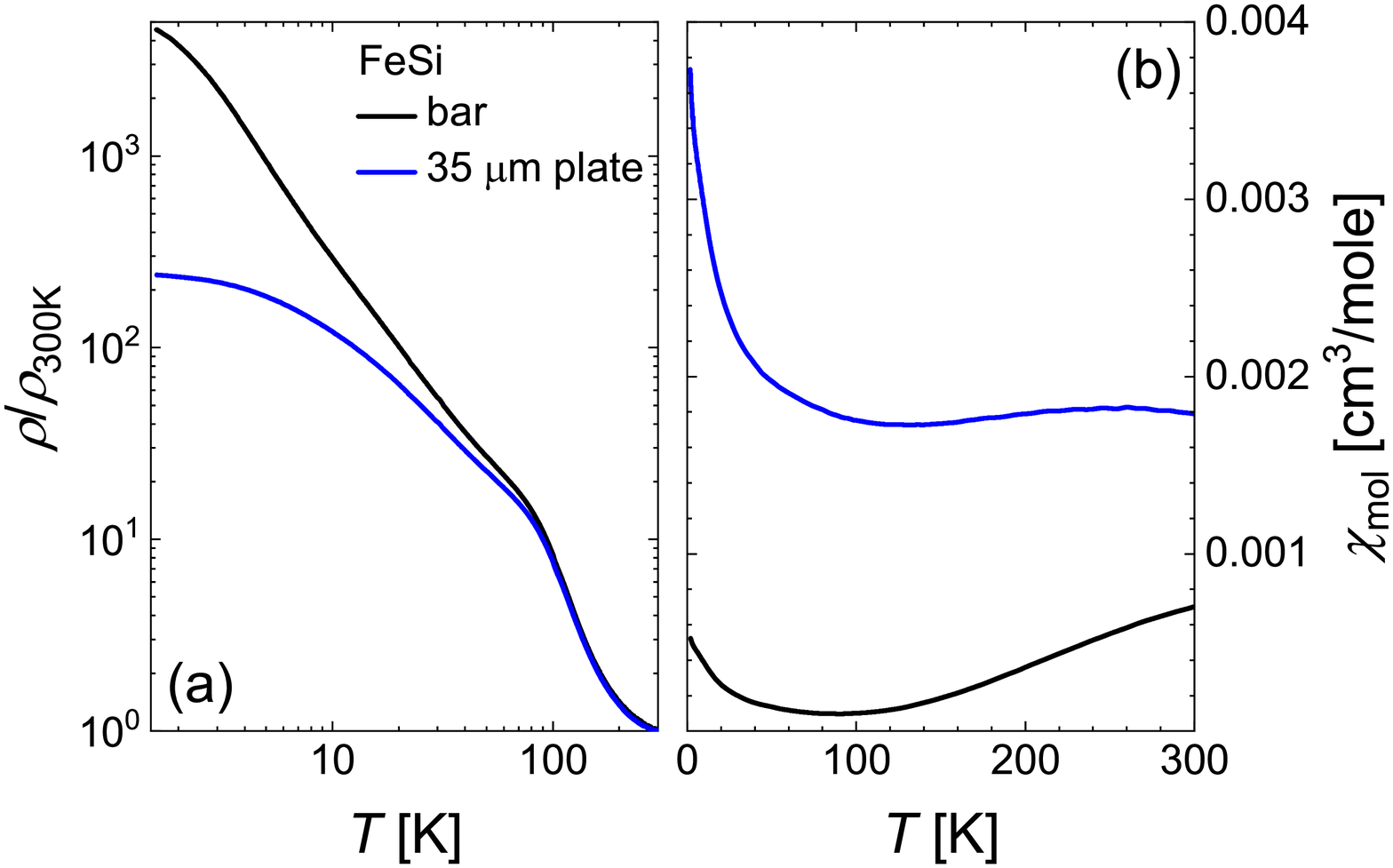}
		\caption{\label{comp} (a) Comparison of the normalized resistivity $\rho / \rho_{300\,\mathrm{K}}$ of single crystalline FeSi measured on a bar-shaped and a thin-plate sample. (b) Susceptibility for the same two samples FeSi; for details see text.}
\end{figure}

In Fig. \ref{comp} (a) we compare these two samples with respect to the normalized resistivity $\rho / \rho_{300\,\mathrm{K}}$ in a log-log-representation. From the figure it is evident that the normalized resistivity of the thin plate deviates from that of the bar-shaped sample below about 100\,K. This finding is qualitatively in line with the observation of Fang et al. \cite{Fang2018} of a metal-to-semiconductor transition in differently sized single crystals FeSi. It verifies the existence of a significant electrical surface conductivity in FeSi, which at low temperatures partially masks the insulating behavior of the bulk of the sample. As we show below, doping with Co in Fe$_{1-x}$Co$_x$Si substantially increases bulk conductivity, and thus reduces the relevance of surface conductivity. Effectively, we find that we can disregard conduction from such surface states in a resistivity measurement at least for doping with Co of more than one percent.

We have also measured the magnetic susceptibility for our two samples FeSi, plotted in Fig. \ref{comp} (b). It has previously been noted that even for single-crystalline FeSi there is always a low-temperature upturn of the susceptibility \cite{Jaccarino1967,Schlesinger1993}. It is usually associated to magnetic (Fe) impurities, although it has been impossible to suppress or diminish this impurity contribution by different preparation techniques. As can be seen, also for our bar-shaped crystal FeSi we observe the typical behavior \cite{Jaccarino1967} with a broad susceptibility maximum above room temperature and the Curie tail at low temperatures. Remarkably, our thin plate sample has a substantially (an order of magnitude at low $T$) increased Curie-like susceptibility background. First, this observation may suggest that polishing the sample damages the surface to the effect that free Fe particles are produced, giving rise to a larger Curie tail. Secondly, and more exotically, these magnetic particles will reside on the surface of the sample, that is in the spatial range of the conducting surface states. It raises the question about the interplay of electronic and magnetic properties in particular at the surface of FeSi, and the possibility that the existence and residual coupling of magnetic moments is associated to a local metallic environment.

\begin{figure}[h]
    \includegraphics[width=1\linewidth]{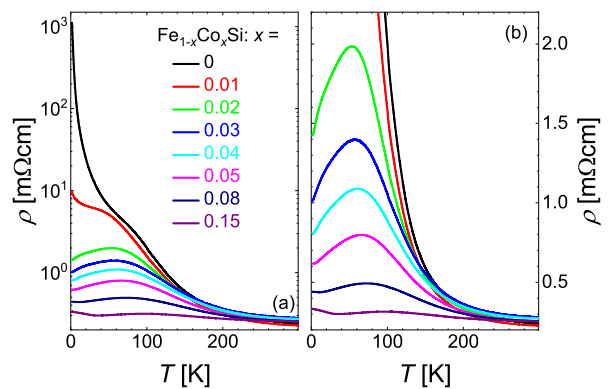}
		\caption{\label{rho} Temperature dependent zero field resistivity $\rho (T)$ of Fe$_{1-x}$Co$_x$Si, $0 \leq x \leq 0.15$, (a) plotted on a logarithmic and (b) linear scale; for details see text.}
\end{figure}

Having thus characterized the relevance of metallic surface states, we proceed with the zero field resistivity of our single-crystalline bar-shaped samples Fe$_{1-x}$Co$_x$Si. In Fig. \ref{rho} we plot the resistivity along the cubic main axis [100] on a logarithmic and linear scale as function of temperature $T$. Globally, the behavior is in full accordance with previous observations: FeSi itself exhibits a gapped behavior, implying it to be in an insulating state at $T = 0$\,K. To quantify the charge gap, we fit the high temperature data $> 200$\,K (to minimize the influence of the surface states) by $\propto \exp \left( \Delta_\mathrm{g} / 2 k_\mathrm{B} T \right)$. This approach yields a gap $\Delta_\mathrm{g} \sim 700$\,K, in good agreement with for instance Ref. \cite{Schlesinger1993} (fit not included in the graph). 

Alloying with Co induces a MIT, apparent from the drastic change of the overall behavior of $\rho$ from insulating to (badly) metallic, with a residual resistivity of $250 \mu\Omega$cm for $x = 0.15$ at lowest temperatures. For $x \geq 0.02$, the low temperature resistivity now has a metallic character $\mathrm{d} \rho / \mathrm{d} T > 0$, while a broad resistive maximum in an intermediate temperature range $\sim 50$\,K has been associated to a remembrance of the narrow gap band structure of FeSi \cite{Lacerda1993,Onose2005} (Fig. \ref{rho}). Notably, from around 200\,K upwards the resistivity for all samples is of similar magnitude, implying that all gap-related features in the resistivity are either overcome by thermal excitations over and/or closing of the gap. 

\begin{figure}[h]
    \includegraphics[width=1\linewidth]{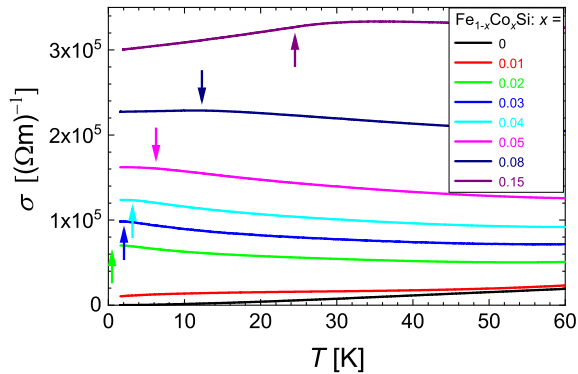}
		\caption{\label{sigma} Temperature dependent zero field conductivity $\sigma (T)$ of Fe$_{1-x}$Co$_x$Si, $0 \leq x \leq 0.15$. Arrows denote magnetic ordering temperatures $T_{\mathrm{HM}}$, for details see text.}
\end{figure}

\begin{figure}[h]
    \includegraphics[width=1\linewidth]{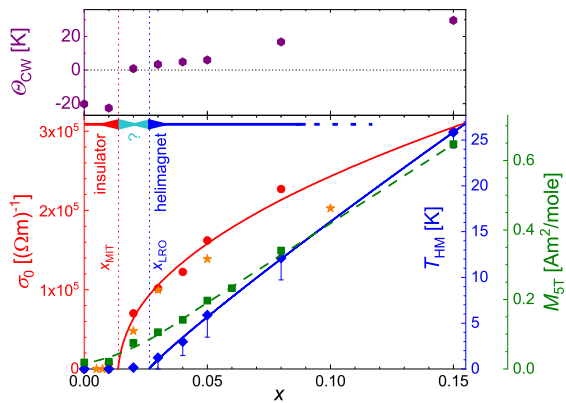}
		\caption{\label{phase} Composition dependence $x$ of the zero temperature conductivity $\sigma_0$ (left scale, red bullets), magnetic ordering temperature $T_{\mathrm{HM}}$ (first right scale, blue diamonds), induced magnetic moment $M_{5\mathrm{T}}$ at 5\,T (second right scale, green squares) and Curie-Weiss temperature $\Theta_{\mathrm{CW}}$ (upper panel) of Fe$_{1-x}$Co$_x$Si, $0 \leq x \leq 0.15$. In the $\sigma_0$-plot we have included the values determined by Chernikov et al. \cite{Chernikov1997} and Manyala et al. \cite{Manyala2000,Manyala2004,Manyala2008} for their polycrystalline samples (orange stars); for details see text.}
\end{figure}

To quantify the MIT we examine the conductivity $\sigma (T) = \rho^{-1} (T)$ plotted for Fe$_{1-x}$Co$_x$Si in Fig. \ref{sigma}. From these data we extract the zero temperature conductivity $\sigma (T \rightarrow 0) = \sigma_0$ presented in Fig. \ref{phase} as function of alloying $x$ (red left scale). This plot visualizes the fundamental change in behavior from insulating to metallic around $x = 0.01$, in agreement with the Refs. \cite{Chernikov1997,Manyala2000,Manyala2004,Manyala2008} ($\sigma_0$--data from these references included in the plot (orange stars)). Increasing the Co concentration beyond the MIT leads to a significant increase of the conductivity, with the absolute value of $\sigma_0$ increasing by an order of magnitude with varying $x$ from 0.01 to 0.02 (Fig. \ref{sigma}).

It was reported \cite{Chernikov1997} that at temperatures $< 1$\,K there is a residual zero-temperature conductivity $\sigma_0 \sim 4$\,$(\Omega$m$)^{-1}$. As we do not cover this temperature range, it might very well also be the case for our crystals. In view of recent studies \cite{Fang2018,Jiang2013} on FeSi and SmB$_6$ and our own findings, it appears to reflect conducting surface states. 

To parametrize the MIT accurately, we draw on the observation of a similarity to classical semiconductors by fitting within critical scaling theory \cite{Lee1985,diTusa1997,Manyala2008} our doping dependence of the zero-temperature conductivity with $\sigma_0 (x) = \sigma(0) (x - x_{\mathrm{MIT}})^{\nu}$. Certainly, the finite metallic surface conductivity might slightly affect the outcome of our fitting procedure. Still, from a fit to the data we obtain values $\sigma(0) = 8.2(1.2) \cdot 10^5$\,$(\Omega$m$)^{-1}$, $x_{\mathrm{MIT}} = 0.014(4)$ and $\nu = 0.50(6)$, consistent with critical scaling theory and similar to previous reports \cite{Chernikov1997,Manyala2000,Manyala2004,Manyala2008} (fit included in the figure as solid red line).
	
\begin{figure}[h]
    \includegraphics[width=1\linewidth]{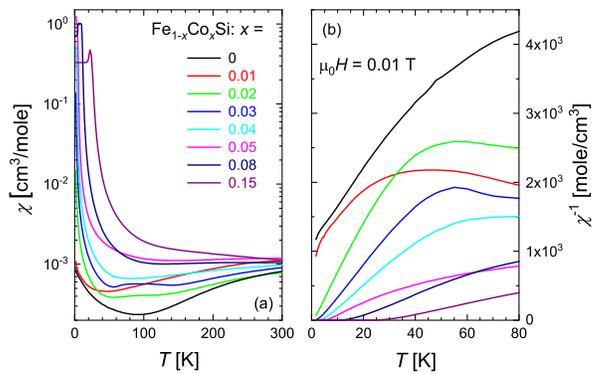}
		\caption{\label{chi} (a) Temperature dependent susceptibility $\chi (T)$, plotted on a logarithmic scale, and (b) inverse susceptibility $\chi^{-1} (T)$ of Fe$_{1-x}$Co$_x$Si, $0 \leq x \leq 0.15$; for details see text.}
\end{figure}

As the next step, knowledge of the underlying magnetic state is necessary. Therefore, in Fig. \ref{chi} we plot the susceptibility $\chi$ (on a logarithmic scale) and inverse susceptibility $\chi^{-1}$ (on a linear scale) of single-crystalline Fe$_{1-x}$Co$_x$Si on different temperature scales measured in 0.01\,T. For some of the samples ($x = 0.01, 0.02, 0.03$) we observe a weak structure in $\chi (T)$ in an intermediate temperature range $\sim 50 - 150$\,K. Zero-field cooled vs. field cooled measurement routines reveal a slight history dependence of these signatures, suggesting that they arise from a small amount of ferro-/ferrimagnetic particles in our samples. Using a toy model for a simple estimate, we might assume that these spurious signals arise for instance from single crystal grain boundaries which might locally produce small grain boundary Fe inclusions. Then, already less than 0.04\,\% of such grain boundary clusters would be sufficient to account for the history dependence of $\chi (T)$. Therefore, these weak additional magnetic signatures are extrinsic and we will not consider them further.

Starting with FeSi, the well-known paramagnetic susceptibility, with a maximum at higher temperatures $\sim 500$\,K is observed, together with a low-temperature Curie-like upturn, so far attributed to a minute amount of magnetic impurities \cite{Schlesinger1993}. Using the argument from Ref. \cite{Schlesinger1993}, the Curie tail would be accounted for by less than 0.2\,\% per formula unit of $S = \frac{3}{2}$ impurity moments. 

The maximum in the susceptibility of FeSi was attributed to an activated behavior across an energy barrier $\Delta_\mathrm{m}$ in the spin excitation spectrum \cite{Schlesinger1993}. For temperatures $T \ll \Delta_\mathrm{m} /k_\mathrm{B}$ it allows to fit the susceptibility by $\chi (T) = \left( C /T \right) \cdot \exp\left(-\Delta_\mathrm{m} /k_\mathrm{B} T\right)$, with $C$ as a constant that in principle measures the spin of the magnetic moments. Accordingly, we can fit the data for FeSi at temperatures above $\sim 150$\,K with a gap of 515\,K, in agreement with Ref. \cite{Schlesinger1993} (not shown).
	
With Co alloying, the high temperature susceptibility maximum broadens and/or shifts to lower temperatures, and has been replaced by an essentially Curie-Weiss-like susceptibility already at 5\,\% Co-doping. If we assume that a shift of the maximum to lower temperatures produces such behavior, a corresponding gap fit applied to Fe$_{0.99}$Co$_{0.01}$Si leads to a substantially reduced gap of $\sim 358$\,K (although the matching with the experiment is substantially worse than for FeSi). It would imply that also with respect to the magnetic properties a minute amount of Co doping is sufficient to suppress gap-like features in the spin excitation spectra.  

At low temperatures all samples exhibit a Curie-Weiss-like upturn of the susceptibility. If we take the approach that this Curie-Weiss-like behavior is associated to magnetic moments with a residual magnetic coupling, then, according to the Curie-Weiss law, the extrapolated intercept of the inverse susceptibility with the temperature axis, the Curie-Weiss temperature $\Theta_{\mathrm{CW}}$, is a measure of the coupling strength. Within this concept, we find all samples Fe$_{1-x}$Co$_x$Si with $x \geq 0.02$ to have a positive $\Theta_{\mathrm{CW}}$, corresponding essentially to a finite ferromagnetic coupling of these samples (see $\chi^{-1}(T)$ in Fig. \ref{chi}). As shown in the purple upper panel of Fig. \ref{phase}, where we display the $x$ dependence of $\Theta_{\mathrm{CW}}$, in this compositional range a linear rise of $\Theta_{\mathrm{CW}}$ with $x$ attests to the strengthening of the magnetic coupling.

Conversely, for FeSi and Fe$_{0.99}$Co$_{0.01}$Si the same construction leaves us with antiferromagnetic Curie-Weiss temperatures $\Theta_{\mathrm{CW}}$ of about -20\,K. As pointed out, these samples have spin excitation gaps much larger than the corresponding values $\Theta_{\mathrm{CW}}$, implying that here we consider diluted magnetic moments in an insulator, {\it i.e.}, a different type of magnetic coupling. Taking these observations together, based on the global behavior of the susceptibility alone, in the metallic regime of the alloying phase diagram we find a finite ferromagnetic coupling, implying that we should observe signatures of long-range magnetic order for these compositions.

To verify this point firmly, we have analyzed the susceptibility and magnetization data for Fe$_{1-x}$Co$_x$Si to extract the ordering temperature $T_{\mathrm{HM}}$ using various approaches as follows: a.) for our susceptibility data $\chi (T)$ we have determined the second temperature derivative, choosing the inflection point as $T_{\mathrm{HM}}$; b.) we have performed a modified Arrott-plot analysis \cite{Arrott1967} to derive $T_{\mathrm{HM}}$; c.) we have parametrized the critical behavior close to $T_{\mathrm{HM}}$ within the framework of the Heisenberg model \cite{Zhang2016}; d.) we have used the Inoue-Shimizu model \cite{Inoue1982} in the generalization by Brommer \cite{Brommer1989} as extension of the Landau-description of phase transitions to establish $T_{\mathrm{HM}}$. 

\begin{figure}[h]
    \includegraphics[width=1\linewidth]{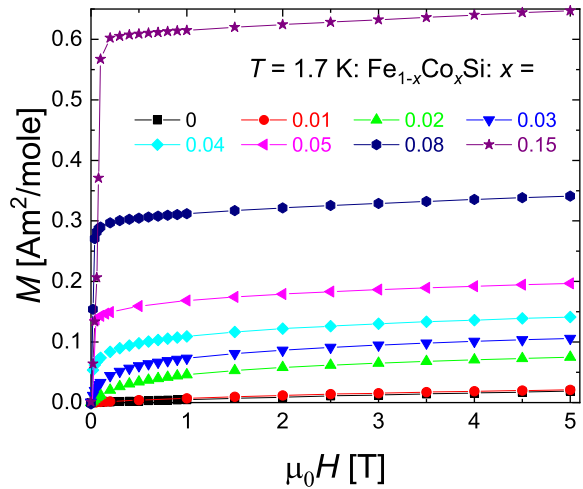}
		\caption{\label{mag} Magnetization $M$ as function of field $H$ measured at 1.7\,K of Fe$_{1-x}$Co$_x$Si, $0 \leq x \leq 0.15$; for details see text.}
\end{figure}

The experimental basis of these analyses are susceptibility data (see above) and magnetization measurements $M (H)$ on our samples Fe$_{1-x}$Co$_x$Si. As an example, in Fig. \ref{mag} we present $M (H)$ at the base temperature of 1.7\,K in fields up to 5\,T. Globally, the figure illustrates the expected behavior: for $x \leq 0.01$ the magnetization is basically flat and close to zero, but for larger $x$ rises continuously and almost linearily with concentration. Given that the helimagnetic order in Fe$_{1-x}$Co$_x$Si is easily field-polarized, the basic field dependence of $M (H)$ for $x \geq 0.02$ is essentially that of a soft ferromagnet. Notably, only in the insulator-to-metal crossover range $0.01 \rightarrow x \rightarrow 0.03$ there is some curvature in the doping evolution of $M (H)$. This is visualized in Fig. \ref{phase}, where we include the doping dependence of the induced magnetic moment $M_{5 \mathrm{T}}$ at 5\,T and 1.7\,K (green outer right scale). 

In the following, we present the different types of analysis and corresponding results for an exemplary case Fe$_{1-x}$Co$_x$Si, $x = 0.05$, with an extended description of the analysis in the supplement \cite{supplement}. Aside from the four approaches we have attempted related types of analysis such as the common Arrott-plot or using different types of criticality (Ising etc.). In result, we find that these approaches either work very badly (Arrott-plot) or do not improve the data parametrization as compared to the approaches discussed here in detail. Overall, the different types of data analysis lead to slightly different ordering temperatures $T_{\mathrm{HM}}$, which we will discuss in more detail below.

\begin{figure}[h]
    \includegraphics[width=1\linewidth]{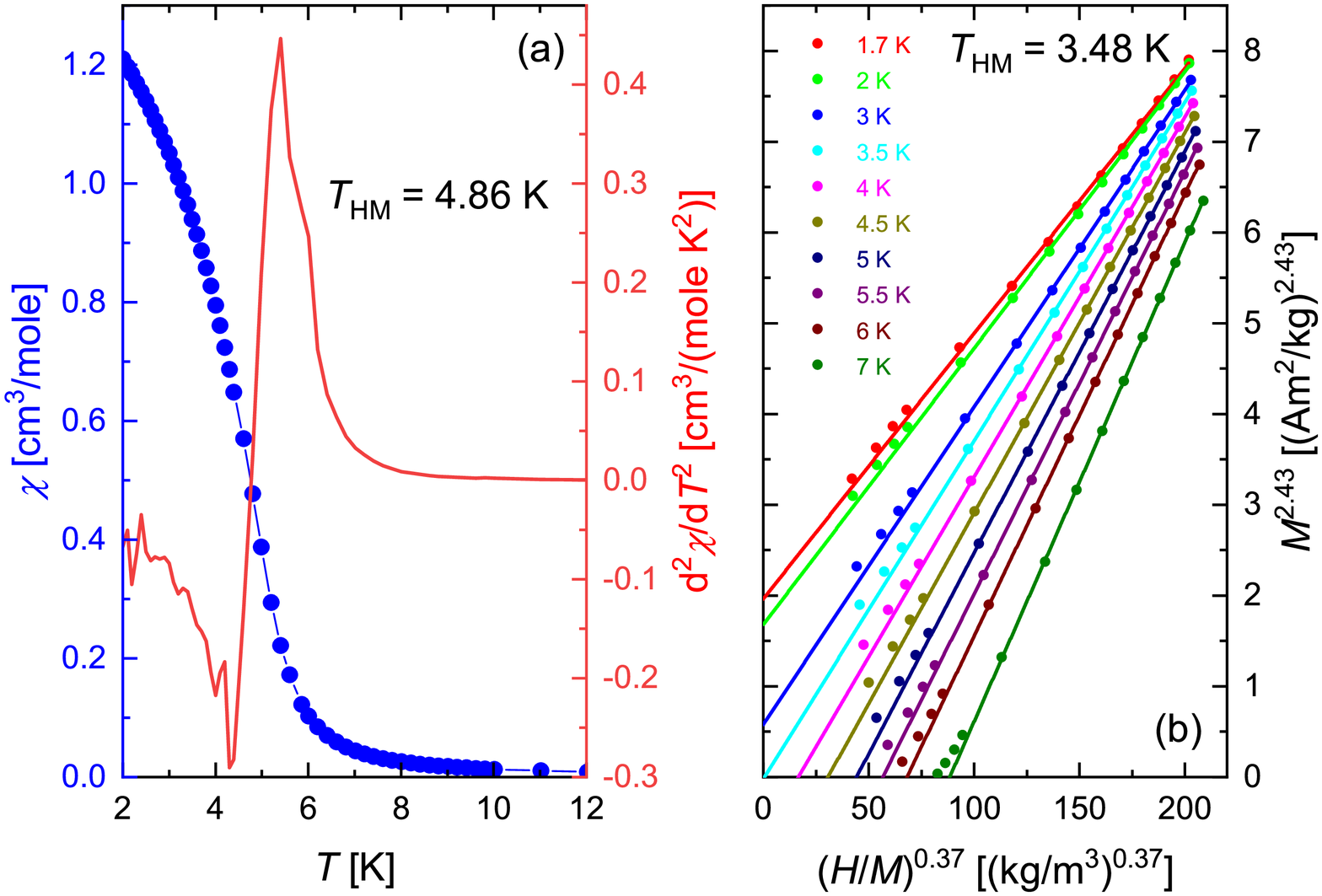}
		\caption{\label{der} (a) Temperature dependent susceptibility $\chi (T)$, measured in $\mu_0 H = 0.01$\,T, and second temperature derivative $\mathrm{d}^2 \chi / \mathrm{d} T^2$ of Fe$_{0.95}$Co$_{0.05}$Si. (b) Modified Arrott-plot analysis of the magnetization of Fe$_{0.95}$Co$_{0.05}$Si plotted as $M^{2.43}$ vs. $(H/M)^{0.37}$; for details see text.}
\end{figure}

We start with the determination of $T_{\mathrm{HM}}$ from the inflection point of the susceptibility $\chi (T)$ for Fe$_{0.95}$Co$_{0.05}$Si (Fig. \ref{der} (a)). The inverse susceptibility $\chi^{-1} (T)$ in a field of $\mu_0 H = 0.01$\,T suggests a magnetic coupling strength somewhat below 10\,K (Fig. \ref{chi}). For low magnetic fields and ignoring demagnetization effects, the inflection point of $\chi (T)$ for a material with a ferri-/ferromagnetic susceptibility signature represents an approximation of the onset of long-range (sublattice) magnetic order. It is derived by numerically calculating the zero intercept of the second derivative $\mathrm{d}^2 \chi (T)/ \mathrm{d} T^2$ included in Fig. \ref{der} (a). This way, from the figure we obtain a transition temperature $T_{\mathrm{HM}}^{\mathrm{susz}} \left( \chi \right) = 4.86$\,K.

Next, as a simple mean-field Arrott-plot analysis does not properly parametrize the experimental data, we have performed a modified Arrott-plot analysis \cite{Arrott1967}. In this approach, by plotting $M^y$ vs. $(H/M)^z$, with $H$ the magnetic field strength, the free parameters $y$ and $z$ are varied to maximize the data range of a linear dependence $M^y = A + B \cdot (H/M)^z$, with $A, B$ the derived free fit parameters. In the spirit of the Arrott-plot analysis, the temperature where $A$ becomes zero is then taken as $T_{\mathrm{HM}}$. It is a phenomenological approach to incorporate the scaling laws of critical phenomena in a comparatively simple data handling procedure. In our case, we find as optimum solution a plot $M^{2.43}$ vs. $(H/M)^{0.37}$ for our magnetization data on Fe$_{0.95}$Co$_{0.05}$Si (Fig. \ref{der} (b)). This in turn leads to an ordering temperature $T_{\mathrm{HM}}^{\mathrm{mod}} = 3.48$\,K.

\begin{figure}[h]
    \includegraphics[width=1\linewidth]{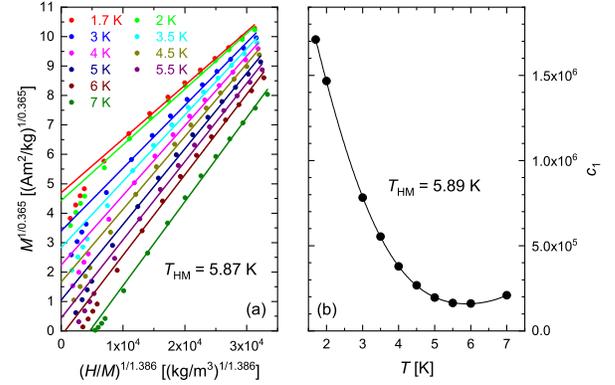}
		\caption{\label{heisen} (a) Determination of the transition temperature $T_{\mathrm{HM}}$ of Fe$_{0.95}$Co$_{0.05}$Si from the magnetization $M (H)$ assuming 3D-Heisenberg criticality. (b) Parameter $c_1$ as function of temperature from an analysis of $M (H)$ using the Inoue-Shimizu model to extract $T_{\mathrm{HM}}$; for details see text.}
\end{figure}

The concept of scaling laws acting close to a critical temperature $T_{\mathrm{HM}}$ is made explicit by observing that $\left( \frac{H}{M} \right)^{\frac{1}{\gamma}} = a \left( \frac{T - T_{\mathrm{HM}}}{T_{\mathrm{HM}}} \right) + b \cdot M^{\frac{1}{\beta}}$, with critical exponents $\gamma, \beta$. Choosing 3D-Heisenberg criticality, we use $\gamma = 1.386$ and $\beta = 0.365$, resulting in a corresponding plot $M^{\frac{1}{\beta}}$ vs. $\left( \frac{H}{M} \right)^{\frac{1}{\gamma}}$ in Fig. \ref{heisen} (a). From this procedure we obtain $T_{\mathrm{HM}}^{\mathrm{Hei}} = 5.87$\,K.

Finally, for the Landau parametrization of magnetic phase transitions the free energy $F$ of a magnetic material is expanded in multiples of the square of the magnetization $M^2$. This approach can be refined for mixed compounds etc. by coupling of the different subsystems \cite{Inoue1982,Brommer1989}. It results in calculating a Lagrange multiplier $\lambda$ from the magnetization derivative of the free energy of the system, $\mathrm{d} F/ \mathrm{d} M = \lambda$, with $\lambda = c_1 M + c_3 M^3 + c_5 M^5$ parametrized using the magnetization. As set out in detail in Ref. \cite{Brommer1989}, the minimization procedure yields the parameters $c_1 (T)$, $c_3 (T)$ and $c_5 (T)$, with the minimum of $c_1 (T)$ defining $T_{\mathrm{HM}}$ and the sign of $c_3 (T)$ detailing the character of the phase transition (1st or 2nd order). We have carried out this analysis for our samples Fe$_{1-x}$Co$_x$Si, with the temperature dependence of $c_1 (T)$ depicted in Fig. \ref{heisen} (b) for the sample $x = 0.05$. From this Inoue-Shimizu analysis we obtain a transition temperature $T_{\mathrm{HM}}^{\mathrm{IS}} = 5.89$\,K.

Combining the results from the different types of analysis for our example Fe$_{0.95}$Co$_{0.05}$Si, we thus obtain a set of transition temperatures $T_{\mathrm{HM}}$ ranging from 3.48\,K to 5.89\,K. For further discussion, we decide to take the highest of the four determined values as ordering temperature $T_{\mathrm{HM}}^*$. For graphic representation in Fig. \ref{phase} (blue inner right scale) we include the alloying dependence $T_{\mathrm{HM}} (x)$ of Fe$_{1-x}$Co$_x$Si in the form of $T_{\mathrm{HM}}^* - \Delta T_{\mathrm{HM}}$, with $\Delta T_{\mathrm{HM}}$ chosen that all values $T_{\mathrm{HM}}$ derived from the different types of analysis are within the error bar (for $x = 0.05$: $\Delta T_{\mathrm{HM}} = 2.41$\,K). Regarding the coefficient $c_3(T)$ specifying if a transition is of 1st or 2nd order, we find that in the extended Inoue-Shimizu model the value of $c_3(T)$ changes sign within the range of the uncertainty $\Delta T_{\mathrm{HM}}$ for all samples. We therefore can not draw a definite conclusion on the nature of the magnetic transition based on the sign of $c_3(T)$.

The dependence $T_{\mathrm{HM}} (x)$ now verifies our observation that for samples Fe$_{1-x}$Co$_x$Si close to $x = 0.02$ long range magnetic order develops. A fit for instance of the transition temperatures with $T_{\mathrm{HM}}^* \propto \left( x - x_{\mathrm{LRO}} \right) ^{\eta}$ yields $x_{\mathrm{LRO}} = 0.026(2)$ and $\eta = 0.92(4)$ (solid blue line in Fig. \ref{phase}). A similar fit, but now taking $T_{\mathrm{HM}}$ as average value of the temperature range $T_{\mathrm{HM}}^* - \Delta T_{\mathrm{HM}}$ yields $x_{\mathrm{LRO}} = 0.031(5)$ and $\eta = 0.92(13)$. Altogether, the analysis results in a close-to-linear ($\eta \sim 0.9 - 1$) concentration $x$ dependence of the ordering temperature $T_{\mathrm{HM}}$ and a critical concentration $x_{\mathrm{LRO}}$ of onset of magnetic order just one percent above the concentration $x_{\mathrm{MIT}}$ of the MIT. Actually, if we consider the error bars of $x_{\mathrm{LRO}}$ and $x_{\mathrm{MIT}}$, the two concentrations almost overlap.

\begin{figure}[h]
    \includegraphics[width=1\linewidth]{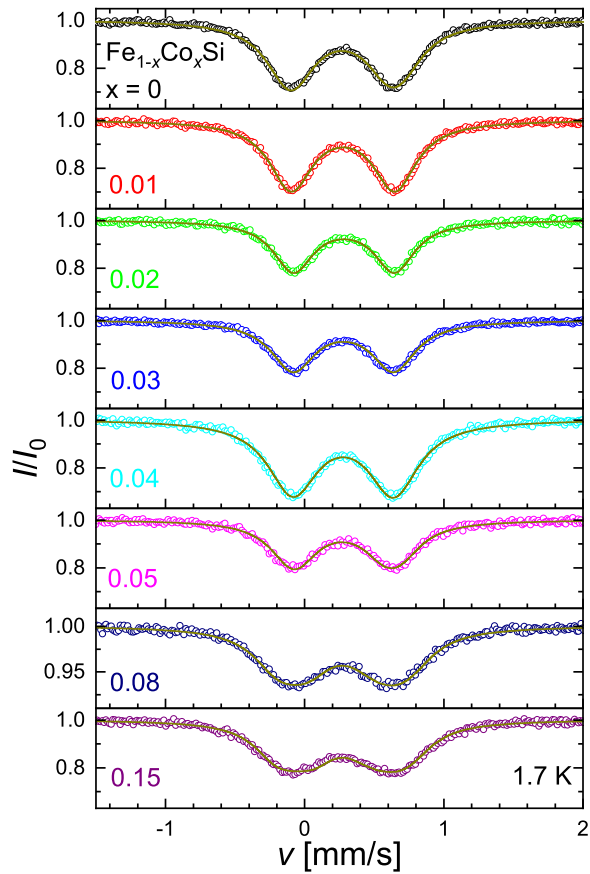}
		\caption{\label{Moess1} M\"ossbauer spectra of single-crystalline Fe$_{1-x}$Co$_x$Si at 1.7\,K; for details see text.}
\end{figure}

As pointed out above, we have prepared thin slices of single-crystalline samples Fe$_{1-x}$Co$_x$Si through polishing with a thickness of a few ten $\mu$m to perform M\"ossbauer spectroscopy as a microscopic probe of the magnetic and electronic properties. In Fig. \ref{Moess1} we present an overview of the experimental results on these samples at a base temperature of 1.7\,K. Here we plot the normalized transmission $I / I_0$, with $I_0$ being the number of counts at large Doppler velocities far away from the absorption lines. Since it was not possible to define the actual sample thickness in the polishing process to a very accurately defined value, there is a slight thickness variation between different samples by up to $\sim 20 \mu$m. This leads to slight variations in the depth of the absorption pattern that can be seen in this plot.
	
Qualitatively, for all samples a symmetric doublet spectrum with a finite isomer shift is detected. For FeSi, in previous M\"ossbauer spectroscopy experiments on polycrystalline powder, this absorption spectrum was attributed to a quadrupole splitting from a non-zero electric field gradient, consistent with the Fe site symmetry in the $B20$-structure \cite{Wertheim1965}. Upon alloying with Co, for small concentrations $x \leq 0.03$ the doublet spectrum persists, while starting with $x = 0.04$ a broadening of the doublet is observed. The broadening stems from helical magnetic order, consistent with the magnetic phase diagram depicted in Fig. \ref{phase}, where at an experimental temperature of 1.7\,K static local magnetic fields at the Fe site ought to be first observable at $x = 0.04$.

\begin{figure}[h]
    \includegraphics[width=1\linewidth]{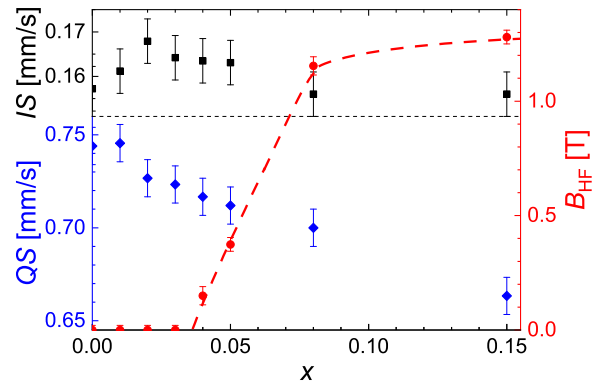}
		\caption{\label{Moess2} Compositional dependence $x$ of $IS$, $QS$ and $B_{\mathrm{HF}}$ for single-crystalline Fe$_{1-x}$Co$_x$Si at 1.7\,K; for details see text.}
\end{figure}

Following the procedure set out in the Refs. \cite{Wertheim1965,Forthaus2011,Dubovtsev1972} we analyze the data starting with FeSi. In our fit we use the free parameters $IS$ and $QS$, giving values $IS = 0.157(5)$\,mm/s, $QS = 0.744(5)$\,mm/s in good agreement with previous reports \cite{Wertheim1965,Forthaus2011,Dubovtsev1972}. The large experimental line width of 0.42\,mm/s is related to the absorber thickness. For the samples $x \geq 0.04$ we observe an additional broadening that is attributed to a magnetic hyperfine field $B_{\mathrm{HF}}$. Strictly speaking, this way we model the helical magnetic state of Fe$_{1-x}$Co$_x$Si, $x \geq 0.04$, as a ferromagnetic one. However, as can be seen from Fig. \ref{Moess1} and \ref{Moess2} this broadening is small and thus the local field distribution in this helical magnet cannot be distinguished in M\"ossbauer spectroscopy from a weak ferromagnetic one (see discussion of a similar situation in NbFe$_2$ \cite{Rauch2015,Willwater2022}). From the fit of the experimental data for all samples (solid lines in Fig. \ref{Moess1}) we obtain the compositional dependence of the fit parameters $IS$, $QS$ and $B_{\mathrm{HF}}$ summarized in Fig. \ref{Moess2}. The general trend of $IS (x)$ and $QS (x)$ is consistent with the findings in previous works \cite{Forthaus2011,Dubovtsev1972}.

The magnetic hyperfine field exhibits an alloying dependence fully consistent with the magnetic phase diagram derived from the bulk magnetic properties (Fig. \ref{phase}). In the temperature range available for the experiment, static magnetic order in the volume of the samples is observable using a microscopic technique for alloying values $x = 0.04$ and above. Consistent with a neutron scattering study on Fe$_{1-x}$Co$_x$Si \cite{Ishimoto1992}, the derived internal magnetic fields in the (sub)-Tesla-range reflect very small ordered magnetic moments $\sim 0.01 - 0.1$\,\textmu$_{\mathrm{B}}$/(Fe/Co)-atom and thus weak magnetic order inherent to the vicinity to a magnetic quantum critical point.

\begin{figure}[h]
    \includegraphics[width=1\linewidth]{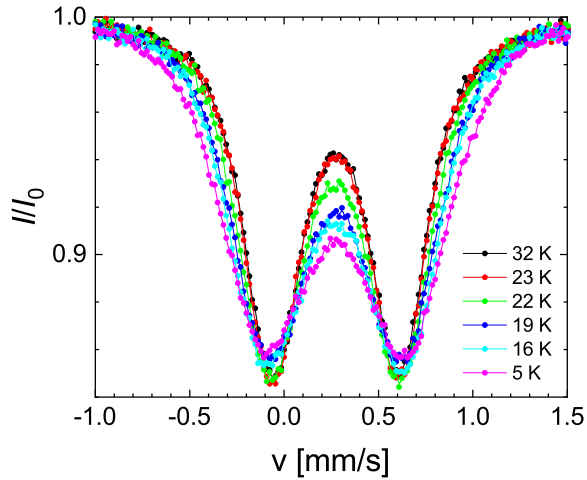}
		\caption{\label{Tdep} Temperature dependent M\"ossbauer spectroscopy data for single-crystalline Fe$_{0,85}$Co$_{0,15}$Si; for details see text.}
\end{figure}

\begin{figure}[h]
    \includegraphics[width=1\linewidth]{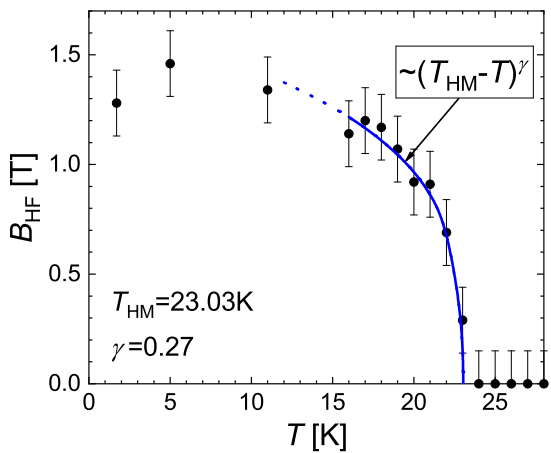}
		\caption{\label{order} Temperature dependence of $B_{\mathrm{HF}}$ of single-crystalline Fe$_{0,85}$Co$_{0,15}$Si. The solid line indicates a critical fit to the data close to $T_{\mathrm{HM}}$; for details see text.}
\end{figure}

Next, we characterize the magnetic order parameter by studying the hyperfine field for selected samples. In Fig. \ref{Tdep} we plot the central ranges of the M\"ossbauer spectra taken for Fe$_{0,85}$Co$_{0,15}$Si as function of temperature. Starting around 23\,K, the doublet spectra broaden due to the onset of magnetic order. Following the above fitting routine, from the data we extract the temperature dependence of $B_{\mathrm{HF}}$ depicted in Fig. \ref{order}. A critical fit to the data close to the ordering temperature $B_{\mathrm{HF}} \propto ( T_{\mathrm{HM}} - T )^\gamma$ yields $T_{\mathrm{HM}} = 23.03(7)$\,K and $\gamma = 0.27(4)$ (solid line in Fig. \ref{order}). Most importantly, the value of $T_{\mathrm{HM}}$ experimentally obtained from the microscopic probe M\"ossbauer spectroscopy is in decent agreement with the values derived from the magnetization/susceptibility analysis ranging from 24.12\,K to 25.83\,K, thus validating the analysis of the bulk magnetic data.

	\section{Discussion}
	
Summarizing our experimental findings for Fe$_{1-x}$Co$_x$Si, we have established a close coincidence of the metal-insulator transition at a composition $x_{\mathrm{MIT}} = 0.014(4)$ and a quantum critical magnetic-non-magnetic transition at $x_{\mathrm{LRO}} \sim 0.026 - 0.031$. If there truly is a difference between $x_{\mathrm{MIT}}$ and $x_{\mathrm{LRO}}$, it would represent a very small section $0.014 < x \lesssim 0.031$ of the phase diagram that constitutes a regime of a low carrier metal with a large magnetic susceptibility and quite unusual physical properties such as the possible formation of magnetic polarons etc. (Fig. \ref{phase}). However, the question that needs to be addressed first is if $x_{\mathrm{MIT}}$ and $x_{\mathrm{LRO}}$ can experimentally be firmly distinguished, {\it i.e.}, the two transitions can be considered to be distinct.

As detailed in Ref. \cite{Willwater2022} in the analysis of the phase diagram for the quantum critical/weakly ferromagnetic system NbFe$_2$, given that both MIT and LRO in Fe$_{1-x}$Co$_x$Si occur in the very dilute Co alloying limit we may assume that a small statistical distribution of local compositions exists in our samples. As set out in Ref. \cite{Willwater2022} for a basic atomic mixing model, for alloying values of $x = 0.014$ and $x = 0.026 - 0.031$ we may expect distributions of local compositions of $\pm ~ 0.002$ and 0.003 around the nominal composition. 

In an experimental study such as ours this occurrence might slightly affect the actually determined compositional value $x$ of any such transition. For instance, for a magnetic/non-magnetic transition a distribution of local compositions will tend to promote short range order at the expense of long-range magnetic order. In other words, the experimentally determined value $x_{LRO}$ will be shifted towards the LRO side of the phase diagram, {\it i.e.}, might be slightly too large in our case. Conversely, for a metal-insulator transition studied by means of conductivity measurements, in samples with a local compositional distribution a percolative metallic conductivity path may form, masking insulating behavior in the bulk of the samples. Therefore, the value $x_{\mathrm{MIT}}$ will be shifted towards the insulating side of the MIT, {\it i.e.}, might be slightly too small in our case. 

Hence, for Fe$_{1-x}$Co$_x$Si a systematic shift of $x_{\mathrm{MIT}}$ to smaller and $x_{\mathrm{LRO}}$ to larger compositional values may occur. Then, correcting for this systematic shift and including the experimental error detailed above we would arrive at values $x_{\mathrm{MIT}} = 0.016(4)$ and $x_{\mathrm{LRO}} \sim 0.023(2) - 0.028(5)$, {\it i.e.}, matching values within error bars. Therefore, for all practical purposes we have to conclude that MIT and LRO critical compositions are probably experimentally not clearly distinguishable. This observation raises the question about the mechanism(s) behind these (joint) MIT/QPT in Fe$_{1-x}$Co$_x$Si, which relates back to the issue of the magnetic character of the small gap semiconductor FeSi. 

Two main concepts have been put forth to account for the magnetic behavior of FeSi. On the one hand, spin fluctuation theory has been invoked to account for the basic magnetic properties \cite{Takahashi1979,Takahashi1983,Takahashi1997,Takahashi1998}. Taken in combination with band structure calculations \cite{Mattheiss1993} it was reported that a single electron picture captures the essential properties of the small gap semiconductor FeSi. Further, extending the band structure calculations by incorporating a Coulomb interaction $U$ revealed an instability of the band structure towards a metallic magnetic state, with the proposal of a field induced MIT to occur in FeSi \cite{Anisimov1996,Neef2006}. 

This concept was further worked out \cite{Anisimov2002,Plencner2009,Dian2014} to account for the properties of the alloying series FeSi$_{1-x}$Ge$_x$. For this series a first-order insulator-to-ferromagnetic metal transition was reported for $x \approx 0.25$, which was interpretetd as result of a tuning of the strength of the Coulomb interaction $U$ with $x$. It reflects the instability of the band structure of FeSi towards a metallic magnetic state noted in Ref. \cite{Anisimov1996}.

On the other hand, the concept of FeSi as a Kondo insulator was proposed, raising the prospect of novel correlation physics apparent in FeSi \cite{Aeppli1992,Schlesinger1993,Fu1994,Mandrus1995,Fu1995}. 
Notably, also the first-order insulator-to-ferromagnetic metal transition was presented within this framework \cite{Yeo2003}. So far, however, direct tests of the Kondo insulator scenario have failed to produce firm evidence for this approach. Only more recently, attempts have been undertaken to merge the different views into a combined picture of correlation effects in a band insulator by using more advanced theoretical tools such as density functional and dynamical mean-field theory \cite{Tomczak2012}.

In the context of these previous observations our findings regarding the electronic and magnetic properties of Fe$_{1-x}$Co$_x$Si stand out in various aspects. At this point, four types \cite{Mani2002} of ''controlled'' doping experiments (that is isoelectronic alloying or changing electron count by one) have been performed on FeSi, that is \cite{Manyala2008,diTusa1997,diTusa1998,Anisimov2002,Yeo2003}: Fe$_{1-x}$Co$_x$Si, Fe$_{1-x}$Mn$_x$Si, FeSi$_{1-x}$Al$_x$ and FeSi$_{1-x}$Ge$_x$. For these series, both Fe$_{1-x}$Mn$_x$Si and FeSi$_{1-x}$Al$_x$ exhibit MITs for lowering alloying levels, while FeSi$_{1-x}$Ge$_x$ transforms in a 1st order transition into a ferromagnetic metal with alloying. 

With respect to the MIT the behavior of Fe$_{1-x}$Co$_x$Si is quite similar to Fe$_{1-x}$Mn$_x$Si and FeSi$_{1-x}$Al$_x$. For all systems the critical concentrations $x_{\mathrm{MIT}}$ for the MITs is in the low percentage range. The $x$-dependence of the zero-temperature conductivity $\sigma_0$ can be parameterized within critical scaling theory - in other words, the MITs appear to behave in a rather common fashion. In contrast, regarding the transition from a non-magnetic to a magnetic state, the behavior of Fe$_{1-x}$Co$_x$Si is in stark contrast to that of FeSi$_{1-x}$Ge$_x$. While regarding the alloying dependence the first series exhibits the typical behavior of quantum criticality, for the latter the transition is discontinuous as function of $x$ and of 1st order nature. Most remarkably, in Fe$_{1-x}$Co$_x$Si the two critical transitions MIT and QPT (almost) coincide regarding their alloying dependence, strongly suggesting a common cause of their appearance. 

More specifically, qualitatively, the $x$ dependence of the QPT clearly bears resemblance to related phenomena in weak ferromagnets close to a magnetic instability \cite{Brando2016} and which conceptually is in principle accounted for by the self-consistent renormalization theory of spin fluctuations \cite{Moriya}. The experimental data suggest that upon approaching the QPT from the LRO side the ordering temperature $T_{\mathrm{HM}}$ and ordered moment $\mu_{\mathrm{ord}}$ vanish/become very small. The unusal aspect is that quantum criticality must occur in the limit of a very small carrier density as result of the MIT. Such behavior may be qualitatively in line with the modeling put forth in Ref. \cite{Tomczak2012}. 

This modeling for FeSi implies that spin and charge response are closely linked, this way reproducing the strong temperature dependence of various physical properties and the concomitant crossover from low temperature insulating to high temperature metallic behavior. Applying this view to our sample series Fe$_{1-x}$Co$_x$Si, Co alloying appears to suppress both spin and charge gap in a similar and quite dramatic fashion. This way, with the closing of the spin and charge gaps the ground state of the system transforms via a QPT into a LRO state. Taken together, with the detailed experimental description of the associated properties presented here, we believe that Fe$_{1-x}$Co$_x$Si lends itself for a thorough microscopic theoretical study of the underlying physical mechanisms, this in particular in the parameter range of a vanishing carrier density.
	
	\begin{acknowledgments}
We acknowledge fruitful discussions with U. K. R\"o\ss ler and J. Aarts.	 
	\end{acknowledgments}

\end{document}